\ifx\pdfoutput\undefined
	\documentclass[dvips,apjl]{emulateapj}
  \DeclareGraphicsExtensions{.eps}
\else
	\documentclass[pdftex,apjl]{emulateapj}
  \DeclareGraphicsExtensions{.pdf}
\fi

\usepackage{apjfonts}

\shorttitle{Exact vs. approximate beaming formulas in GRB afterglows}
\shortauthors{C.L. Bianco \& R. Ruffini}

\begin{document}

\title{Exact versus approximate beaming formulas in Gamma-Ray Burst afterglows}

\author{Carlo Luciano Bianco\altaffilmark{1} and Remo Ruffini\altaffilmark{2}}
\affil{ICRANet and ICRA, Piazzale della Repubblica 10, I--65100 Pescara, Italy.}
\affil{Dipartimento di Fisica, Universit\`a di Roma ``La Sapienza'', Piazzale Aldo Moro 5, I-00185 Roma, Italy.}

\altaffiltext{1}{E-mail: bianco@icra.it}
\altaffiltext{2}{E-mail: ruffini@icra.it}

\begin{abstract}
We present the exact analytic expressions to compute, assuming the emitted Gamma-Ray Burst (GRB) radiation is not spherically symmetric but is confined into a narrow jet, the value of the detector arrival time at which we start to ``see'' the sides of the jet, both in the fully radiative and adiabatic regimes. We obtain this result using our exact analytic expressions for the EQuiTemporal Surfaces (EQTSs) in GRB afterglows. We re-examine the validity of three different approximate formulas currently adopted for the adiabatic regime in the GRB literature. We also present an empirical fit of the numerical solutions of the exact equations, compared and contrasted with the three above approximate formulas. The extent of the differences is such as to require a reassessment on the existence and entity of beaming in the cases considered in the current literature, as well as on its consequences on the GRB energetics.
\end{abstract}


\section{Introduction}\label{intro}

After the work by \citet{my94}, the possibility that Gamma-Ray Bursts (GRBs) originate from a beamed emission has been one of the most debated issue about the nature of the GRB sources in the current literature \citep[see e.g.][and references therein]{p04,m06}. In particular, on the ground of the theoretical considerations by \citet{sph99}, it was conjectured that, within the framework of a conical jet model, one may find that the gamma-ray energy released in all GRBs is narrowly clustered around $5 \times 10^{50}$ ergs \citep{fa01}.

In a recent letter \citep{PowerLaws} we analyzed the approximate power-law relations between the Lorentz gamma factor and the radial coordinate usually adopted in the current GRB literature. We pointed out how such relations are found to be mathematically correct but only approximately valid in a very limited range of the physical and astrophysical parameters and in an asymptotic regime which is reached only for a very short time, if any. Therefore, such relations were there shown to be not applicable to GRBs. Instead, the exact analytic solutions of the equations of motion of a relativistic thin and uniform shell expanding in the interstellar medium (ISM) in the fully radiative and adiabatic regimes were there presented.

This program of identifying the exact analytic solutions instead of approximate power-law solution is in this letter carried one step forward. Using the above exact solutions, we here introduce the exact analytic expressions of the relations between the detector arrival time $t_a^d$ of the GRB afterglow radiation and the corresponding half-opening angle $\vartheta$ of the expanding source visible area due to the relativistic beaming \citep[see e.g.][]{Brasile}. Such visible area must be computed not over the spherical surface of the shell, but over the EQuiTemporal Surface (EQTS) of detector arrival time $t_a^d$, i.e. over the surface locus of points which are source of the radiation reaching the observer at the same arrival time $t_a^d$ \citep[see][for details]{EQTS_ApJL,EQTS_ApJL2}. The exact analytic expressions for the EQTSs in GRB afterglows, which have been presented in \citet{EQTS_ApJL2}, are therefore crucial in our present derivation. This approach clearly differs from the ones in the current literature, which usually neglect the contributions of the radiation emitted from the entire EQTS.

The analytic relations between $t_a^d$ and $\vartheta$ presented in this Letter allow to compute, assuming that the expanding shell is not spherically symmetric but is confined into a narrow jet with half-opening angle $\vartheta_\circ$, the value $(t_a^d)_{jet}$ of the detector arrival time at which we start to ``see'' the sides of the jet. A corresponding ``break'' in the observed light curve should occur later than $(t_a^d)_{jet}$ \citep[see e.g.][]{sph99}. In the current literature, $(t_a^d)_{jet}$ is usually defined as the detector arrival time at which $\gamma \sim 1/\vartheta_\circ$, where $\gamma$ is the Lorentz factor of the expanding shell \citep[see e.g.][and also our Eq.(\ref{Sist}) below]{sph99}. In our formulation we do not consider effects of lateral spreadings of the jet.

In the current literature, in the case of adiabatic regime, different approximate power-law relations between $(t_a^d)_{jet}$ and $\vartheta_\circ$ have been presented, in contrast to each other \citep[see e.g.][]{sph99,pm99,p06}. We show here that in four specific cases of GRBs, encompassing more than $5$ orders of magnitude in energy and more than $2$ orders of magnitude in ISM density, both the one by \citet{pm99} and the one by \citet{sph99} overestimate the exact analytic result. A third relation just presented by \citet{p06} slightly underestimate the exact analytic result. We also present an empirical fit of the numerical solutions of the exact equations for the adiabatic regime, compared and contrasted with the three above approximate relations. In the fully radiative regime, and therefore in the general case, no simple power-law relation of the kind found in the adiabatic regime can be established and the general approach we have outlined has to be followed.

Although evidence for spherically symmetric emission in GRBs is emerging from observations \citep{sa06} and from theoretical argumentations \citep{Spectr1,cospar04}, it is appropriate to develop here an exact theoretical treatment of the relation between $(t_a^d)_{jet}$ and $\vartheta_\circ$. This will allow to make an assessment on the existence and, in the positive case, on the extent of beaming in GRBs, which in turn is going to be essential for establishing their correct energetics.

\section{Analytic formulas for the beaming angle}

The boundary of the visible region of a relativistic thin and uniform shell expanding in the ISM is defined by \citep[see e.g.][and references therein]{Brasile}:
\begin{equation}
\cos\vartheta = \frac{v}{c}\, ,
\label{bound}
\end{equation}
where $\vartheta$ is the angle between the line of sight and the radial expansion velocity of a point on the shell surface, $v$ is the velocity of the expanding shell and $c$ is the speed of light. To find the value of the half-opening beaming angle $\vartheta_\circ$ corresponding to an observed arrival time $(t_a^d)_{jet}$, this equation must be solved together with the equation describing the EQTS of arrival time $(t_a^d)_{jet}$ \citep{EQTS_ApJL2}. In other words, we must solve the following system:
\begin{equation}
\left\{
\begin{array}{rcl}
\cos\vartheta_\circ & = & \frac{v\left(r\right)}{c}\\
\cos\vartheta_\circ & = & \cos\left\{\left.\vartheta \left[r;(t_a^d)_{jet}\right]\right|_{EQTS\left[(t_a^d)_{jet}\right]}\right\}
\end{array}
\right.\, .
\label{Sist}
\end{equation}
It should be noted that, in the limit $\vartheta_\circ \to 0$ and $v \to c$, this definition of $(t_a^d)_{jet}$ is equivalent to the one usually adopted in the current literature (see sec. \ref{intro}).

\subsection{The fully radiative regime}

In this case, the analytic solution of the equations of motion gives \citep[see][]{EQTS_ApJL2,PowerLaws}:
\begin{equation}
\frac{v}{c} = \frac{\sqrt{\left(1-\gamma_\circ^{-2}\right)\left[1+\left(M_{ism}/M_B\right)+\left(M_{ism}/M_B\right)^2\right]}}{1+\left(M_{ism}/M_B\right)\left(1+\gamma_\circ^{-1}\right)\left[1+\textstyle\frac{1}{2}\left(M_{ism}/M_B\right)\right]}\, ,
\label{vRad}
\end{equation}
where $\gamma_\circ$ and $M_B$ are respectively the values of the Lorentz gamma factor and of the mass of the accelerated baryons at the beginning of the afterglow phase and $M_{ism}$ is the value of the ISM matter swept up to radius $r$: $M_\mathrm{ism}=(4\pi/3)m_pn_{ism}(r^3-{r_\circ}^3)$, where $r_\circ$ is the starting radius of the baryonic matter shell, $m_p$ is the proton mass and $n_{ism}$ is the ISM number density. Using the analytic expression for the EQTS given in \citet{EQTS_ApJL2}, Eq.(\ref{Sist}) takes the form:
\begin{equation}
\left\{
\begin{array}{rcl}
\cos\vartheta_\circ & = & \frac{\sqrt{\left(1-\gamma_\circ^{-2}\right)\left[1+\left(M_{ism}/M_B\right)+\left(M_{ism}/M_B\right)^2\right]}}{1+\left(M_{ism}/M_B\right)\left(1+\gamma_\circ^{-1}\right)\left[1+\textstyle\frac{1}{2}\left(M_{ism}/M_B\right)\right]}\\[18pt]
\cos\vartheta_\circ & = & \frac{M_B  - m_i^\circ}{2r\sqrt{C}}\left( {r - r_\circ } \right) +\frac{m_i^\circ r_\circ }{8r\sqrt{C}}\left[ {\left( {\frac{r}{{r_\circ }}} \right)^4  - 1} \right] \\[6pt]
& + & \frac{{r_\circ \sqrt{C} }}{{12rm_i^\circ A^2 }} \ln \left\{ {\frac{{\left[ {A + \left(r/r_\circ\right)} \right]^3 \left(A^3  + 1\right)}}{{\left[A^3  + \left( r/r_\circ \right)^3\right] \left( {A + 1} \right)^3}}} \right\} \\[6pt]
& + & \frac{ct_\circ}{r} - \frac{c(t_a^d)_{jet}}{r\left(1+z\right)} + \frac{r^\star}{r} \\[6pt]
& + & \frac{{r_\circ \sqrt{3C} }}{{6rm_i^\circ A^2 }} \left[ \arctan \frac{{2\left(r/r_\circ\right) - A}}{{A\sqrt{3} }} - \arctan \frac{{2 - A}}{{A\sqrt{3} }}\right]
\end{array}
\right.
\label{SistRad}
\end{equation}
where $t_\circ$ is the value of the time $t$ at the beginning of the afterglow phase, $m_i^\circ=(4/3)\pi m_p n_{\mathrm{ism}} r_\circ^3$, $r^\star$ is the initial size of the expanding source, $A=[(M_B-m_i^\circ)/m_i^\circ]^{1/3}$, $C={M_B}^2(\gamma_\circ-1)/(\gamma_\circ +1)$ and $z$ is the cosmological redshift of the source.

\subsection{The adiabatic regime}

In this case, the analytic solution of the equations of motion gives \citep[see][]{EQTS_ApJL2,PowerLaws}:
\begin{equation}
\frac{v}{c} = \sqrt{\gamma_\circ^2-1}\left(\gamma_\circ+\frac{M_{ism}}{M_B}\right)^{-1}
\label{vAd}
\end{equation}
Using the analytic expression for the EQTS given in \citet{EQTS_ApJL2}, Eq.(\ref{Sist}) takes the form:
\begin{equation}
\left\{
\begin{array}{rcl}
\cos\vartheta_\circ & = & \sqrt{\gamma_\circ^2-1}\left(\gamma_\circ+\frac{M_{ism}}{M_B}\right)^{-1} \\[18pt]
\cos\vartheta_\circ & = & \frac{m_i^\circ}{4M_B\sqrt{\gamma_\circ^2-1}}\left[\left(\frac{r}{r_\circ}\right)^3  - \frac{r_\circ}{r}\right] + \frac{ct_\circ}{r} \\[6pt]
& - & \frac{c(t_a^d)_{jet}}{r\left(1+z\right)} + \frac{r^\star}{r} - \frac{\gamma_\circ-\left(m_i^\circ/M_B\right)}{\sqrt{\gamma_\circ^2-1}}\left[\frac{r_\circ}{r} - 1\right]
\end{array}
\right.
\label{SistAd}
\end{equation}
where all the quantities have the same definition as in Eq.(\ref{SistRad}).

\subsection{The comparison between the two solutions}

\begin{figure}
\includegraphics[width=\hsize,clip]{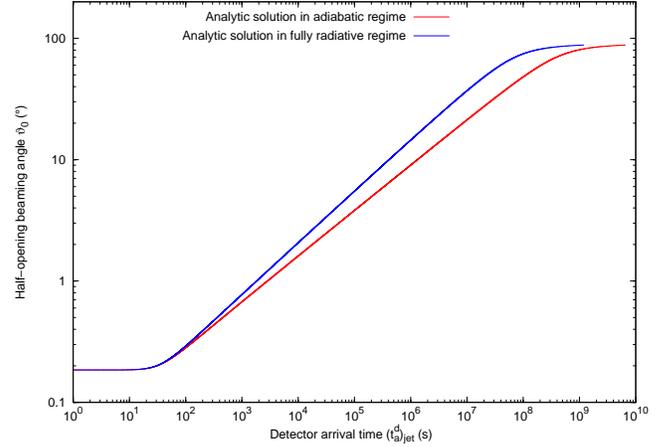}
\caption{Comparison between the numerical solution of Eq.(\ref{SistRad}) assuming fully radiative regime (blue line) and the corresponding one of Eq.(\ref{SistAd}) assuming adiabatic regime (red line). The departure from power-law behavior at small arrival time follows from the constant Lorentz $\gamma$ factor regime, while the one at large angles follows from the approach to the non-relativistic regime \citep[see details in section \ref{fit} and Fig. \ref{Beam_Comp_Num_Fit_Log}, as well as in][]{PowerLaws}.}
\label{Beam_Comp_Rad_Ad}
\end{figure}

In Fig. \ref{Beam_Comp_Rad_Ad} we plot the numerical solutions of both Eq.(\ref{SistRad}), corresponding to the fully radiative regime, and Eq.(\ref{SistAd}), corresponding to the adiabatic one. Both curves have been plotted assuming the same initial conditions, namely the ones of GRB 991216 \citep[see][]{Brasile}.

\section{Comparison with the existing literature}

\begin{figure*}
\includegraphics[width=0.5\hsize,clip]{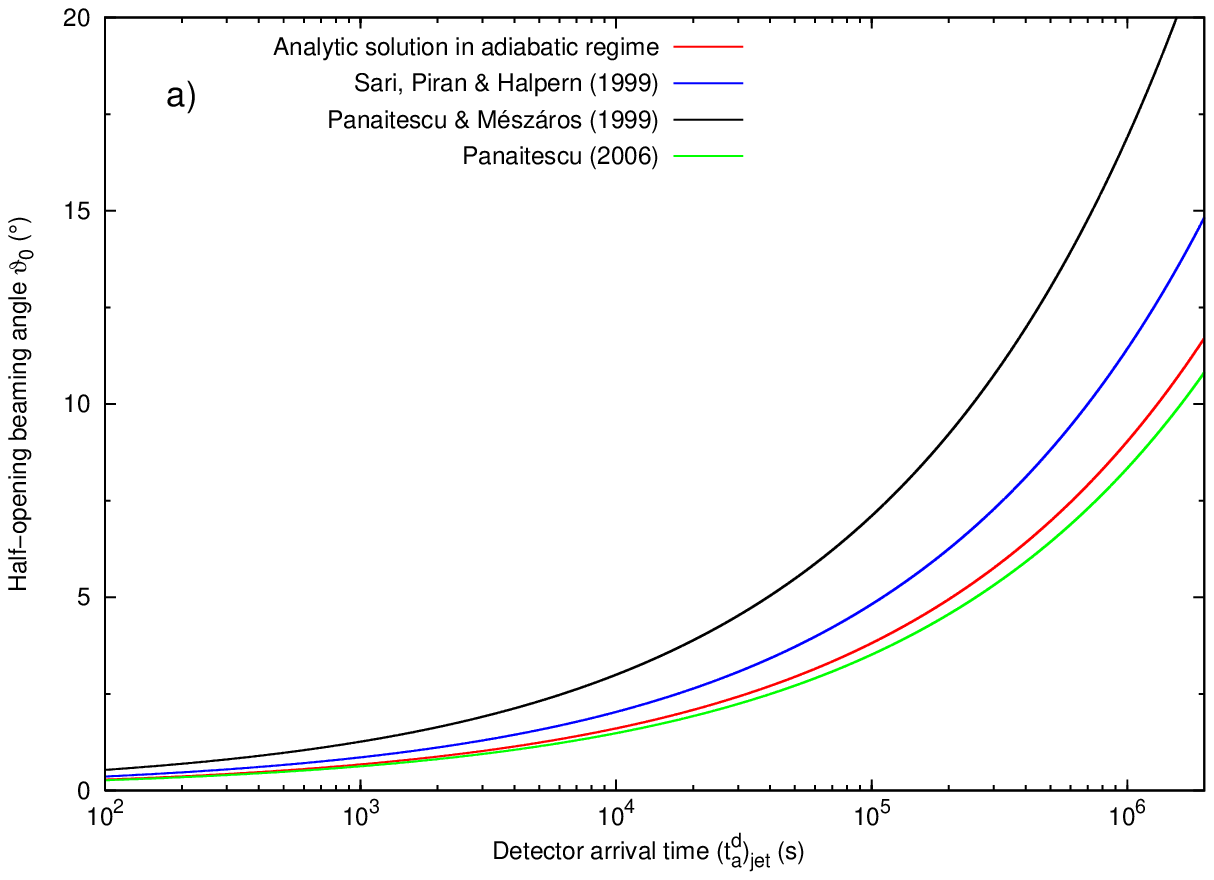}
\includegraphics[width=0.5\hsize,clip]{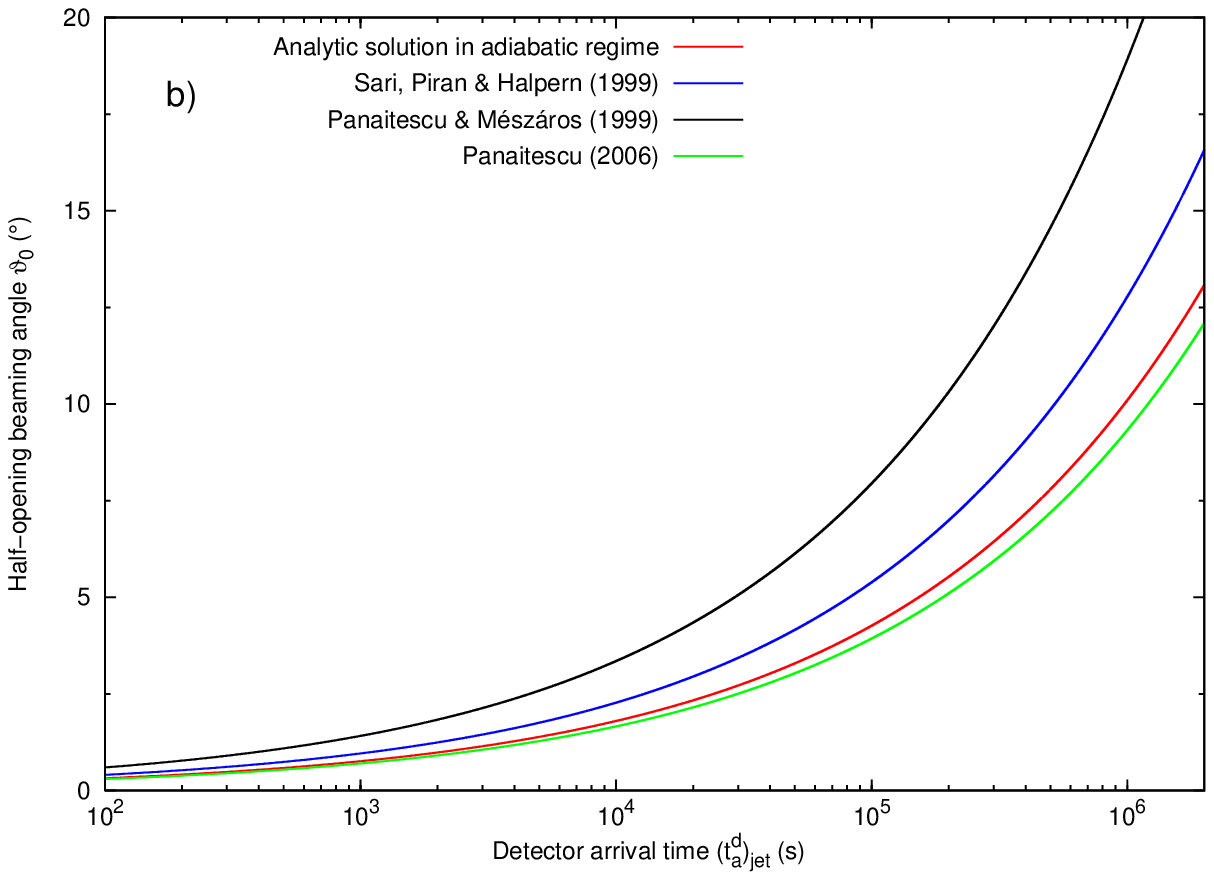}\\
\includegraphics[width=0.5\hsize,clip]{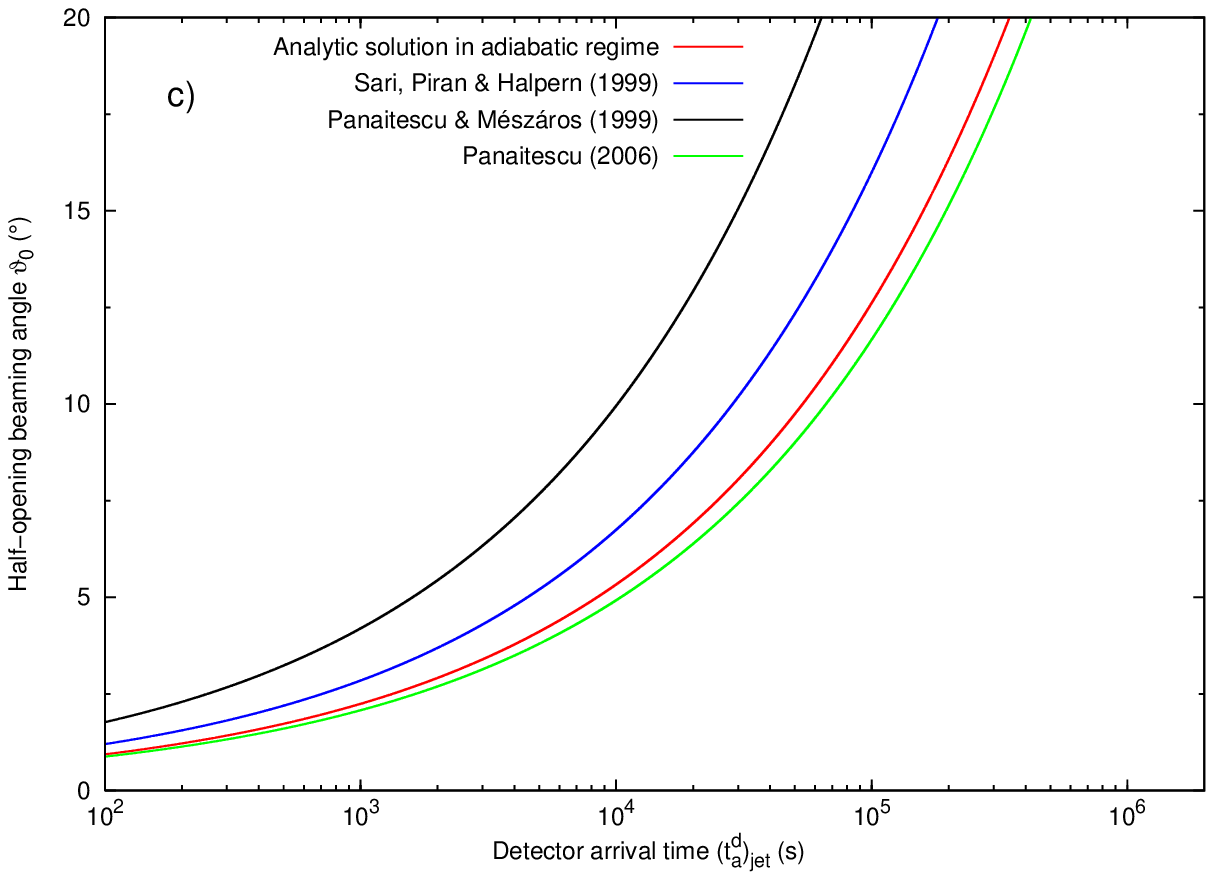}
\includegraphics[width=0.5\hsize,clip]{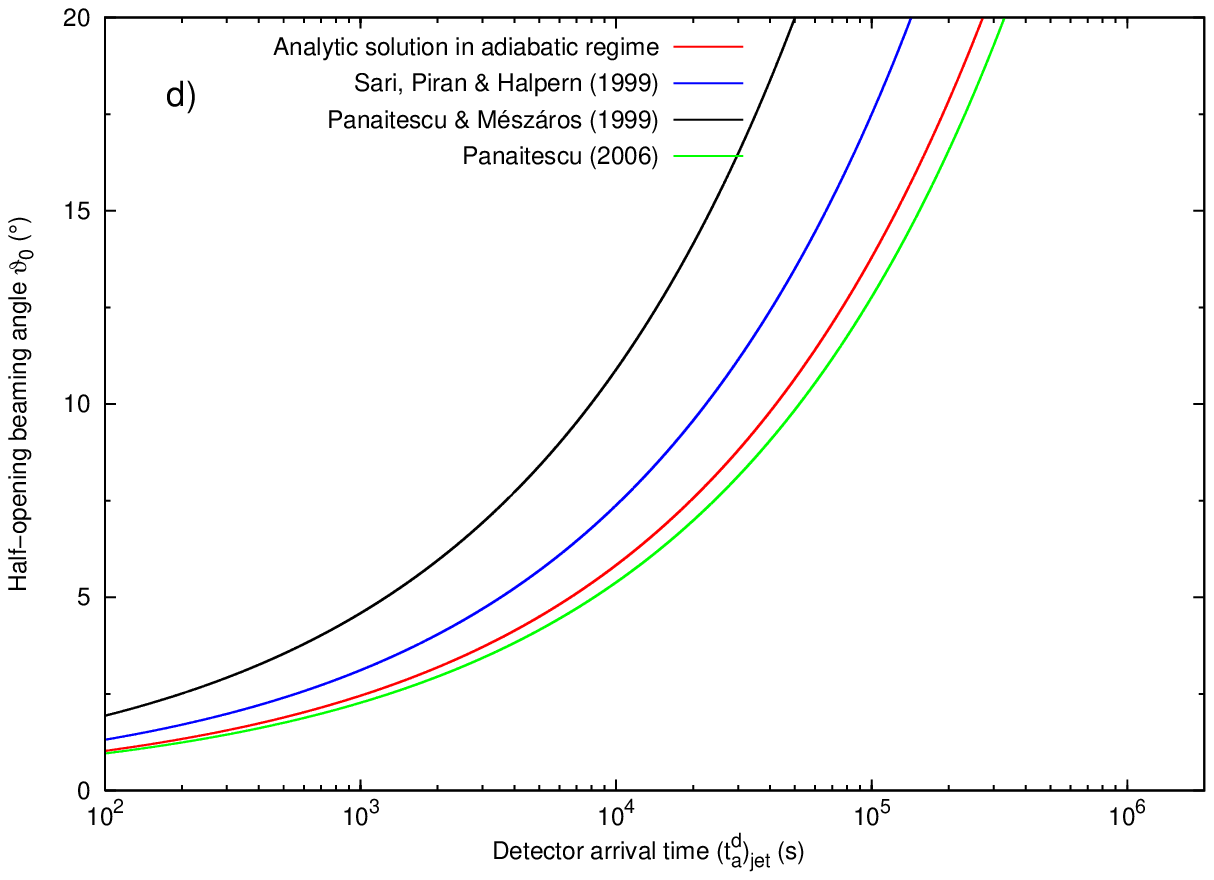}
\caption{Comparison between the numerical solution of Eq.(\ref{SistAd}) (red line) and the corresponding approximate formulas given in Eq.(\ref{ThetaSPH99}) (blue line), in Eq.(\ref{ThetaPM99}) (black line), and in Eq.(\ref{ThetaP06}) (green line). All four curves have been plotted for four different GRBs: a) GRB 991216 \citep[see][]{Brasile}, b) GRB 980519 \citep[see][]{980519}, c) GRB 031203 \citep[see][]{031203}, d) GRB 980425 \citep[see][]{cospar02}. The ranges of the two axes have been chosen to focus on the sole domains of application of the approximate treatments in the current literature.}
\label{Beam_Comp}
\end{figure*}

Three different approximate formulas for the relation between $(t_a^d)_{jet}$ and $\vartheta_\circ$ have been given in the current literature, all assuming the adiabatic regime. \citet{pm99} proposed:
\begin{equation}
\cos\vartheta_\circ \simeq 1 - 5.9\times10^7 \left(\frac{n_{ism}}{E}\right)^{1/4} \left[\frac{(t_a^d)_{jet}}{1+z}\right]^{3/4} \, ,
\label{ThetaPM99}
\end{equation}
\citet{sph99}, instead, advanced:
\begin{equation}
\vartheta_\circ \simeq 7.4\times 10^3 \left(\frac{n_{ism}}{E}\right)^{1/8} \left[\frac{(t_a^d)_{jet}}{1+z}\right]^{3/8} \, .
\label{ThetaSPH99}
\end{equation}
In both Eq.(\ref{ThetaPM99}) and Eq.(\ref{ThetaSPH99}), $(t_a^d)_{jet}$ is measured in seconds, $E$ is the source initial energy measured in ergs and $n_{ism}$ is the ISM number density in particles/cm$^3$. The formula by \citet{sph99} has been applied quite often in the current literature \citep[see e.g.][]{fa01,ggl04,fa05}.

Both Eq.(\ref{ThetaPM99}) and Eq.(\ref{ThetaSPH99}) compute the arrival time of the photons at the detector assuming that all the radiation is emitted at $\vartheta=0$ (i.e. on the line of sight), neglecting the full shape of the EQTSs. Recently, a new expression has been proposed by \citet{p06}, again neglecting the full shape of the EQTSs but assuming that all the radiation is emitted from $\vartheta = 1/\gamma$, i.e. from the boundary of the visible region. Such an expression is:
\begin{equation}
\vartheta_\circ \simeq 5.4\times 10^3 \left(\frac{n_{ism}}{E}\right)^{1/8} \left[\frac{(t_a^d)_{jet}}{1+z}\right]^{3/8} \, .
\label{ThetaP06}
\end{equation}

In Fig. \ref{Beam_Comp} we plot Eq.(\ref{ThetaPM99}), Eq.(\ref{ThetaSPH99}) and Eq.(\ref{ThetaP06}) together with the numerical solution of Eq.(\ref{SistAd}) relative to the adiabatic regime. All four curves have been plotted assuming the same initial conditions for four different GRBs, encompassing more than $5$ orders of magnitude in energy and more than $2$ orders of magnitude in ISM density: a) GRB 991216 \citep[see][]{Brasile}, b) GRB 980519 \citep[see][]{980519}, c) GRB 031203 \citep[see][]{031203}, d) GRB 980425 \citep[see][]{cospar02}. The approximate Eq.(\ref{ThetaSPH99}) by \citet{sph99} and Eq.(\ref{ThetaP06}) by \citet{p06} both imply a power-law relation between $\vartheta_\circ$ and $(t_a^d)_{jet}$ with constant index $3/8$ for any value of $\vartheta_\circ$, while Eq.(\ref{ThetaPM99}) by \citet{pm99} implies a power-law relation with constant index $3/8$ only for $\vartheta_\circ \to 0$ (for greater $\vartheta_\circ$ values the relation is trigonometric).

All the above three approximate treatments are based on the approximate power-law solutions of the GRB afterglow dynamics which have been shown in \citet{PowerLaws} to be not applicable to GRBs. They also do not take fully into account the structure of the EQTSs, although in different ways. Both Eq.(\ref{ThetaPM99}) and Eq.(\ref{ThetaSPH99}), which assume all the radiation coming from $\vartheta=0$, overestimate the behavior of the exact solution. On the other hand, Eq.(\ref{ThetaP06}), which assumes all the radiation coming from $\vartheta\sim 1/\gamma$, is a better approximation than the previous two, but still slightly underestimates the exact solution.

\section{An empirical fit of the numerical solution}\label{fit}

\begin{figure}
\includegraphics[width=\hsize,clip]{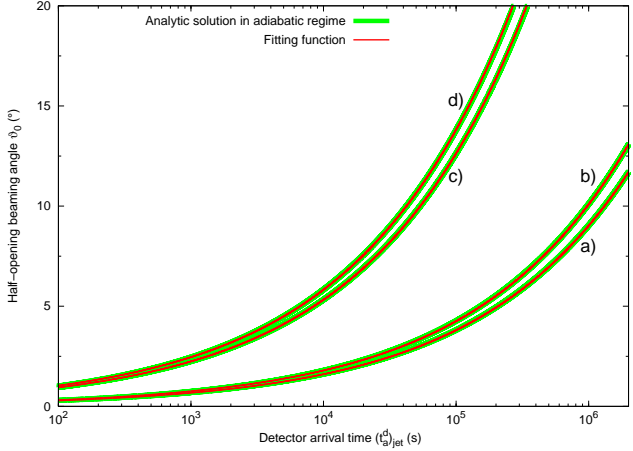}
\caption{The overlapping between the numerical solution of Eq.(\ref{SistAd}) (thick green lines) and the approximate fitting function given in Eq.(\ref{ThetaFIT}) (thin red lines) is shown in the four cases (a--d) represented in Fig. \ref{Beam_Comp}.}
\label{Beam_Comp_Num_Fit}
\end{figure}

\begin{figure}
\includegraphics[width=\hsize,clip]{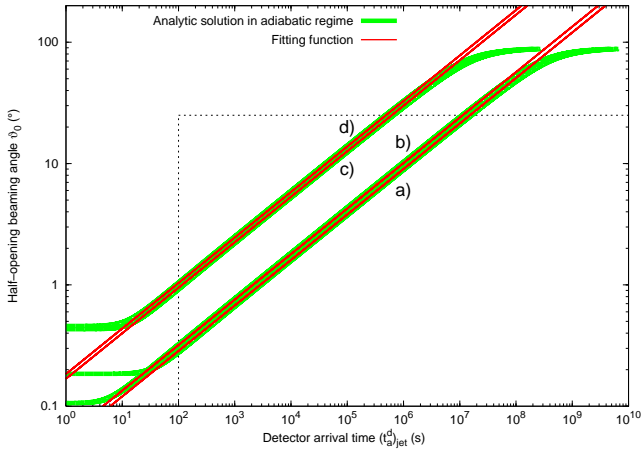}
\caption{Comparison between the numerical solution of Eq.(\ref{SistAd}) (think green lines) and the approximate fitting function given in Eq.(\ref{ThetaFIT}) (thin red lines) in all the four cases (a--d) represented in Fig. \ref{Beam_Comp}. The ranges of the two axes have been chosen to have their full extension (i.e. the one of Fig. \ref{Beam_Comp_Rad_Ad}). The dashed gray lines are the boundaries of the region where the empirical fitting function can be applied.}
\label{Beam_Comp_Num_Fit_Log}
\end{figure}

For completeness, we now fit our exact solution with a suitable explicit functional form in the four cases considered in Fig. \ref{Beam_Comp}. We chose the same functional form of Eq.(\ref{ThetaP06}), which is the closer one to the numerical solution, using the numerical factor in front of it (i.e. $5.4\times 10^3$) as the fitting parameter. We find that the following approximate expression:
\begin{equation}
\vartheta_\circ \simeq 5.84\times 10^3 \left(\frac{n_{ism}}{E}\right)^{1/8} \left[\frac{(t_a^d)_{jet}}{1+z}\right]^{3/8}
\label{ThetaFIT}
\end{equation}
is in agreement with the numerical solution in all the four cases presented in Fig. \ref{Beam_Comp} (see Fig. \ref{Beam_Comp_Num_Fit}). However, if we enlarge the axis ranges to their full extension (i.e. the one of Fig. \ref{Beam_Comp_Rad_Ad}), we see that such approximate empirical fitting formula can only be applied for $\vartheta_\circ < 25^\circ$ \emph{and} $(t_a^d)_{jet} > 10^2$ s (see the gray dashed rectangle in Fig. \ref{Beam_Comp_Num_Fit_Log}).

An equivalent empirical fit in the fully radiative regime is not possible. In this case, indeed, there is a domain in the $((t_a^d)_{jet},\vartheta_\circ)$ plane where the numerical solution shows a power-law dependence on time, with an index $\sim 0.423$ (see Fig. \ref{Beam_Comp_Rad_Ad}). However, the dependence on the energy cannot be factorized out with a simple power-law. Therefore, in the fully radiative regime, which is the relevant one for our GRB model \citep[see e.g.][]{Brasile}, the application of the full Eq.(\ref{SistRad}) does not appear to be avoidable.

\section{Conclusions}

We have presented in Eqs.(\ref{SistRad},\ref{SistAd}) the exact analytic relations between the jet half-opening angle $\vartheta_\circ$ and the detector arrival time $(t_a^d)_{jet}$ at which we start to ``see'' the sides of the jet, which may be used in GRB sources in which an achromatic light curve break is observed. The limiting cases of fully radiative and adiabatic regimes have been outlined. Such relations differs from the approximate ones presented in the current literature in the adiabatic regime: both the ones by \citet{pm99} and by \citet{sph99} overestimate the exact analytic result, while the one just presented by \citet{p06} slightly underestimate it.

For a limited domain in the $((t_a^d)_{jet},\vartheta_\circ)$ plane defined in Fig. \ref{Beam_Comp_Num_Fit_Log}, and only in the adiabatic regime, an empirical fit of the numerical solution of the exact Eq.(\ref{SistAd}) has been given in Eq.(\ref{ThetaFIT}). However, in the fully radiative regime such a simple empirical power-law fit does not exist and the application of the exact Eq.(\ref{SistRad}) is needed. This same situation is expected also to occur in the general case.

In light of the above results, the assertion that the gamma-ray energy released in all GRBs is narrowly clustered around $5 \times 10^{50}$ ergs \citep{fa01} should be reconsidered. In addition, the high quality data by Swift, going without gaps from the ``prompt emission'' all the way to latest afterglow phases, will help in uniquely identifying the equations of motion of the GRB sources and the emission regimes. Consequently, on the ground of the results presented in this Letter, which encompass the different dynamical and emission regimes in GRB afterglow, an assessment on the existence and, in the positive case, on the extent of beaming in GRBs will be possible. This is a step in the determination of their energetics.

\acknowledgments
We thank an anonymous referee for his/her interesting suggestions.

\end{document}